# Anomalous Hall effect in insulating $Ga_{1-x}Mn_xAs$

Sh. U. Yuldashev, H. C. Jeon, H. S. Im and T. W. Kang

Quantum Functional Semiconductor Research Center, Dongguk University, 3-26 Pil-dong, Chung-ku, Seoul 100-715, Korea

S. H. Lee

Department of Physics, Korea University, Seoul Korea, 136-701

J. K. Furdyna

Department of Physics, University of Notre Dame, Notre Dame, Indiana 46556

We have investigated the effect of doping by Te on the anomalous Hall effect in $Ga_{1-x}Mn_xAs$ ($x = 0.085$). For this relatively high value of x the temperature dependence of resistivity shows an insulating behavior. It is well known that in $Ga_{1-x}Mn_xAs$ the Mn ions naturally act as acceptors. Additional doping by Te donors decreases the Curie temperature and increases the anomalous Hall resistivity. With increasing Te concentration the long-range ferromagnetic order in $Ga_{1-x}Mn_xAs$ eventually disappears, and paramagnetic–to–spin glass transition is observed instead. The critical concentration of holes required for establishing ferromagnetic order in the $Ga_{0.915}Mn_{0.085}As$ has been estimated by using the magnetic polaron percolation theory proposed by Kaminski and Das Sarma [Phys. Rev. Lett. **88**, 247202 (2002)].

PACS numbers: 75.50.Pp, 75.70.Ak, 73.50.Jt, 73.61.Ey.

At present, the origin of the anomalous Hall effect in ferromagnetic semiconductors is under active debate.[1,2] Recently, the Berry phase mechanism of AHE in ferromagnetic semiconductors has been proposed by Jungwirth et al.[3] following the initial idea given by Karplus and Luttinger [4] that the AHE in ferromagnets is caused by intrinsic spin-orbit interaction. $Ga_{1-x}Mn_xAs$ is the most thoroughly investigated "workhorse" material used for the study of ferromagnetic properties of diluted magnetic semiconductors. In this context magneto-transport measurements – and especially the anomalous Hall effect -- have been central in elucidating the essential role of free carriers (holes) in establishing the long-range ferromagnetic order in $Ga_{1-x}Mn_xAs$.[3]

While most effort in this area has so far been focused on the AHE in $Ga_{1-x}Mn_xAs$ with a high concentration of free carriers,[5,6] relatively little attention has been given to the case of low concentration of free holes, i.e., when this material shows insulating behavior. In this paper we present the results of Hall measurements on $Ga_{1-x}Mn_xAs$ ( x = 0.085) which as-grown (i.e., without additional doping) shows insulating type of conductivity. To further decrease the hole concentration in this material by compensation, additional counter-doping by Te (a donor impurity) has been carried out.

The $Ga_{1-x}Mn_xAs$ films were grown on semi-insulating (001) GaAs substrates in a Riber 32 MBE system. Prior to $Ga_{1-x}Mn_xAs$ deposition a 100 nm GaAs buffer layer was grown at $590^0C$. The substrate was then cooled to 250 $^0C$ for the growth of a 100 nm low temperature (LT) GaAs buffer, followed by a 300nm layer of $Ga_{1-x}Mn_xAs$ additionally doped by Te. The concentration of Te was controlled by the temperature of the Te source. A $Ga_{1-x}Mn_xAs$ epitaxial layer without Te doping was also grown under identical conditions to serve as a reference sample. The Mn concentration in the layers was estimated from X-ray diffraction measurements and it was additionally confirmed by X-ray microanalysis. A value of x = 0.085 was obtained for all samples in this growth series. The resistivity and Hall effect measurements at different temperatures and magnetic fields were carried out by using a He cryostat equipped with a superconducting coil magnet.

Figure 1 shows the temperature dependence of the resistivity for undoped (A) and Te-doped (B,C) $Ga_{1-x}Mn_xAs$ (x = 0.085) samples. The resistivity curves corresponding to samples A and B showed a local maximum (a shoulder) near their respective Curie temperature $T_C$ (shown by arrows in Fig. 1). This critical behavior of resistivity is commonly observed in ferromagnetic metals and ferromagnetic semiconductors, and is ascribed to the scattering of carriers by magnetic spin fluctuation via exchange interaction.[7] In our previous paper [8] the resistivity maximum near the Curie temperature was successfully discussed in terms of the magnetoimpurity scattering model proposed by Nagaev.[9]

The resistivity curve of sample C does not show the local maximum in its temperature dependence. As seen from Fig.1, the resistivities of samples B and C are much higher in comparison with that of sample A, especially at low temperatures. This result shows that additional doping by Te dramatically reduces the concentration of free holes in $Ga_{1-x}Mn_xAs$. The insert of Fig.1 shows a plot of ln $\rho$ as a function of $1/T^{1/4}$ for sample C. It is seen that all experimental points lie rather well on the straight line. This behavior is consistent with conductance via a Mott variable-range hopping,[10] indicating that the holes are localized in this sample even at room temperature.

Ferromagnetic materials are characterized by a Hall resistivity $\rho_{xy}$ that consists of two components, one associated with the Lorentz force (the normal Hall resistivity), and the other related to the effect of the spin-orbit interaction on the spin-polarized carriers (the anomalous Hall effect). The value of the latter component may exceed the normal Hall contribution by several orders of magnitude.[11] The Hall resistivity can thus be expressed as

$$\rho_{xy} = R_0 B + R_S M \quad , \qquad (1)$$

where $R_0$ and $R_S$ are the normal and the anomalous Hall coefficients, respectively; B is magnetic field; and M is the magnetization. The anomalous Hall coefficient itself depends on the longitudinal resistivity as $R_S \sim \rho_{xx}^n$, where n = 1 or 2 in the case of skew- scattering and side-jump scattering, respectively.[11] On the other hand, the intrinsic (Berry phase) theory[3] of the AHE in ferromagnetic semiconductors has also predicted an n = 2 dependence, but the anomalous Hall coefficient does not depend on the concentration of scattering centers.

Figures 2a-2c show the magnetic field dependence of the Hall resistivity measured at different temperatures for samples A, B, and C, respectively. Prior to the Hall measurements the samples were cooled at zero magnetic field following a proper zero-field-cooled (ZFC) procedure. For all samples investigated the Hall resistivity at low temperatures shows a clearly non-linear dependence on the magnetic field, thus indicating that the measured Hall resistivity is dominated by of the anomalous Hall effect in all three samples. More specifically, it is seen from Figs. 2a-2c that the Hall resistivity for samples A and B decreases with increasing temperature in the entire temperature range examined, while for sample C the Hall resistivity initially increases (between 10K and 30K), and subsequently decreases with further increase of temperature. The increase of Hall resistivity with increasing temperature T is unusual in $Ga_{1-x}Mn_xAs$, because it signals an increase with T of one of the two terms on the right side of Eq.(1). The first term (the ordinary Hall coefficient) for insulating semiconductors usually decreases with increasing T because the concentration of free carriers in these materials increases with T, causing a decrease of resistivity, as clearly seen in the temperature dependence of the resistivity shown in Fig.1. The anomalous Hall coefficient, i.e., the second term in Eq.(1), also decreases with increasing temperature, because $R_S \sim \rho_{xx}^n$. The increase of the Hall resistivity for sample C with increasing temperature observed between 10K and 30K might therefore be connected with increasing magnetization with T in that temperature region. To study this, the temperature dependence of magnetization M(T) for sample C was measured under zero-field-cooled (ZFC) and field-cooled (FC) conditions, as shown in Fig.3. A magnetic field of 0.02T perpendicular to the film plane was used for both of the ZFC and FC magnetization measurements.

It is seen in Fig. 3 that for ZFC conditions the magnetization of sample C decreases with decreasing temperature below 30K. Such behavior is typical for a spin-disordered (spin glass) state in diluted magnetic semiconductors.[12] This result thus shows that in sample C the concentration of holes is insufficient for establishing long-range ferromagnetic order for the entire volume of the sample; and the M(T) curve for this sample displays the paramagnetic–to–spin glass transition at 30K.

We have used the bound magnetic polaron percolation theory proposed by Kaminski and Das Sarma[13,14] to estimate the critical concentration of holes which is necessary for establishing long-range ferromagnetic ordering in $Ga_{1-x}Mn_xAs$ (x = 0.085). According to this theory the effective radius of the bound magnetic polaron is [14]

$$r_{pol}(T) = \frac{a_B}{2} \ln \frac{AJ_0(a_B^3 n_i)^{1/2}}{T}, \quad (2)$$

where A ~ 1, $a_B$ is the Bohr radius of the hole, $J_0$ is the exchange energy between the magnetic ion and the localized hole, and $n_i$ is the concentration of magnetic ions. According to Eq. (2) the effective radius of the magnetic polaron increases with decreasing temperature, so that overlapping of magnetic polaron wave functions can occur. At the critical value of $r_{pol}(n_h)^{1/3} \approx$ 0.86, where $n_h$ is the concentration of holes, the infinite cluster (percolation) of bound magnetic polarons is then expected to spread across the entire sample, thus establishing long-range ferromagnetic order. Using Eq.(2) and the critical parameter mentioned above, we have estimated the critical concentration of holes for $Ga_{1-x}Mn_xAs$ (x = 0.085) at T = 10K to be $n_h \approx$ 3 × 10$^{19}$ cm$^{-3}$. In this calculation we used $a_B$ = 1.39 nm (the Bohr radius of heavy hole in GaAs) and the exchange energy $J_0 \approx J_{pd}/a_B^3$ [13], where $J_{pd} = 50 \pm 5$ meV nm$^3$ is the strength of the exchange coupling between magnetic ions and the free holes in the valence band of Ga$_{1-x}$Mn$_x$As.[3]

At high temperatures (e.g., 300K), the dependence of $\rho_{xy}$ on the magnetic field is linear for all samples (see Fig.2). This linear behavior allows us to *roughly* estimate the free carrier concentration in our samples, even though the anomalous Hall effect contribution to $\rho_{xy}$ may still be substantial at these high temperatures[6]. The resistivity and the hole concentration obtained from Hall measurements at 300K for all investigated samples are summarized in Table I. These results show that the concentration of holes in sample C is indeed lower than the critical value of $3 \times 10^{19}$ cm$^{-3}$ calculated above. It is therefore not surprising that sample C does not show long-range ferromagnetic order at any temperature, but instead shows a spin glass state below 30K. The hole concentration in sample B shown in Table I is also lower than the calculated value of the critical hole concentration, but one must remember that this critical hole concentration was obtained using very simple approximations, therefore it should be considered as rough estimate.

In conclusion, the temperature dependence of the resistivity and the Hall effect measurements was systematically measured on Ga$_{1-x}$Mn$_x$As (x = 0.085) doped by Te. The temperature dependence of the resistivity for all samples investigated shows an insulating behavior. It was observed that the additional Te doping decreases the Curie temperature and increases the resistivity, as would be expected due to compensation by Te donors. At low temperatures the Hall resistivity dependence on magnetic field shows a nonlinear behavior for all samples investigated, indicative of the dominance of AHE in these samples. A paramagnetic–to–spin glass transition was observed at 30K in the magnetization for the Ga$_{1-x}$Mn$_x$As (x = 0.085) specimen with the lowest concentration of holes in this sample series. The spin glass state was observed to cause an unusual behavior of the anomalous Hall resistivity as a function of temperature in this sample below of 30K. More generally, this last sample serves to illustrate that the anomalous Hall effect can be very pronounced even in specimens which do not exhibit long-range ferromagnetic order.

We thank Tomasz Dietl for helpful discussion on the physics of the anomalous Hall effect. This work was supported by the Korea Science and Engineering Foundation (KOSEF) through the Quantum Functional Semiconductor Research Center (QSRC) at Dongguk University; by the KIST Vision 21 project, by the DARPA SpinS Program; and by the National Science Foundation Grant DMR02-45227

**Figure captions**

Fig.1. Temperature dependence of the resistivity for undoped (A) and Te-doped (B,C) samples of $Ga_{1-x}Mn_xAs$ (x = 0.085) at zero magnetic field. Arrows mark the critical temperature Tc deduced from magnetization. In the inset, the plot of $\ln \rho$ vs. $1/T^{1/4}$ is shown for sample C.

Fig.2. Magnetic field dependence of the Hall resistivity at different temperatures for (a) undoped and (b,c) Te-doped samples of $Ga_{1-x}Mn_xAs$ (x = 0.085).

Fig.3. Temperature dependence of magnetization measured at zero-field-cooled (ZFC) and field-cooled (FC) conditions for sample C.

Table I. Hall data for undoped (A) and Te-doped (B and C) samples of $Ga_{1-x}Mn_xAs$ (x = 0.085) measured at 300K.

| Sample | A | B | C |
|---|---|---|---|
| Resistivity, $\Omega$ cm | 0.074 | 0.156 | 0.261 |
| Hole concentration, $cm^{-3}$ | $3.8 \times 10^{19}$ | $1.25 \times 10^{19}$ | $3.4 \times 10^{18}$ |

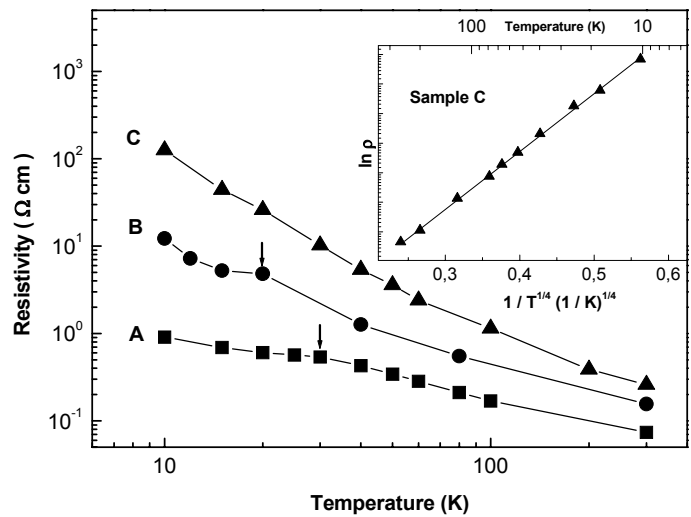

Fig.1. Sh.U.Yuldashev *et al.*

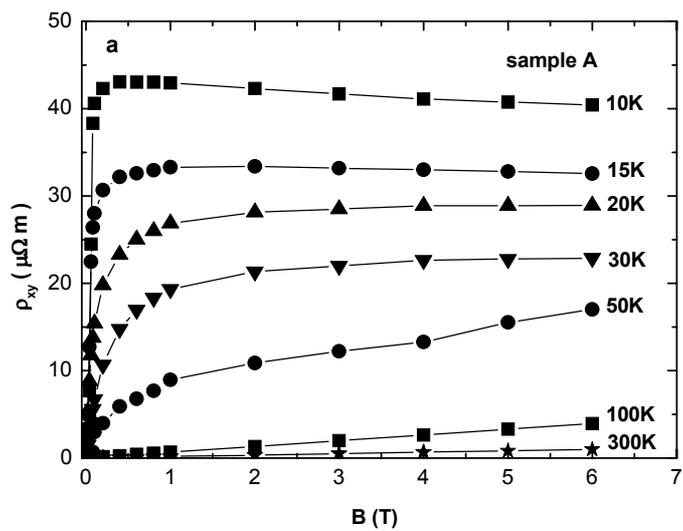

Fig. 2a. Sh.U. Yuldashev *et al.*

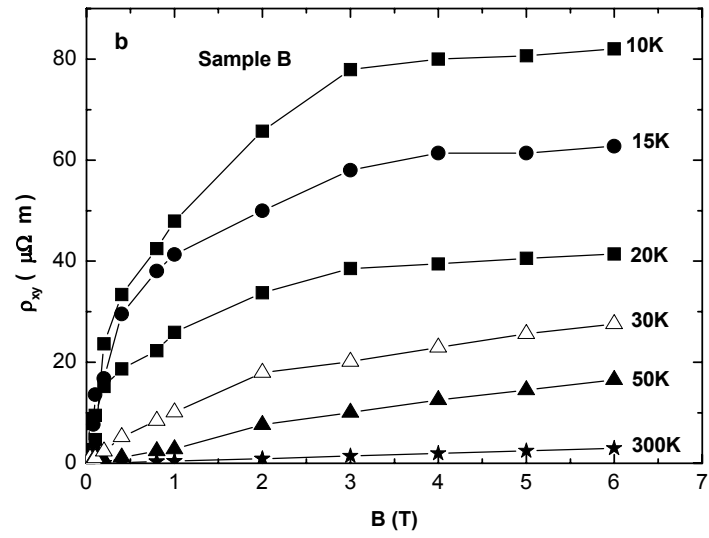

Fig. 2b. Sh.U. Yuldashev *et al.*

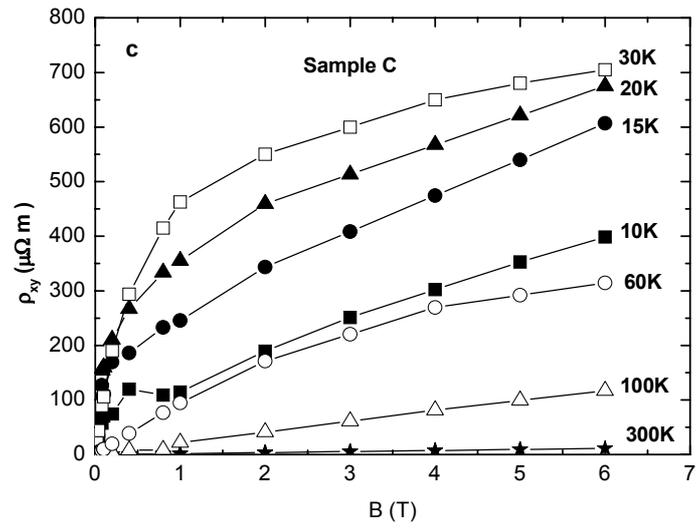

Fig. 2c. Sh.U. Yuldashev *et al.*

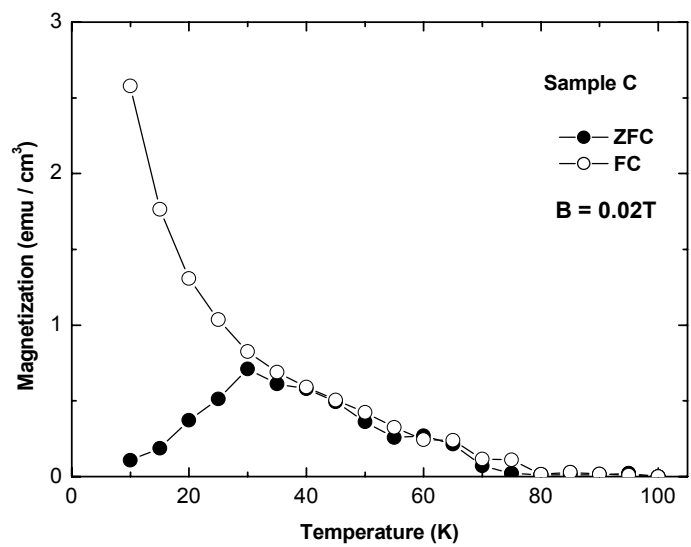

Fig. 3. Sh.U. Yuldashev *et al.*